\UseRawInputEncoding
\documentclass[12pt]{article}
\usepackage[cmtip,arrow]{xy}
\usepackage{pb-diagram,pb-xy}
\usepackage{amsmath}
\usepackage[psamsfonts]{amssymb}
\usepackage{cmmib57}
\usepackage{amsmath,amssymb,amscd}
\usepackage{tikz-cd}
\newcommand{\bPf}{\par\vspace*{-4pt}\indent{\sc Proof.}\enskip}
\newcommand{\ePf}{\medskip}
\def\QED{\hskip0.1em\hfill\null\ \null\nobreak\hfill\kern3pt\vbox{\hrule\hbox
   {\vrule\kern1pt\vbox{\kern1.7pt\hbox{$\scriptscriptstyle{QED}$}
    \kern0.2pt}\kern1pt\vrule}\hrule}}

\def\END{\hskip0.1em\hfill\null\ \null\nobreak\hfill\kern3pt\vbox{\hrule\hbox
   {\vrule\kern1pt\vbox{\kern1.7pt\hbox{$\,\,\,\vspace{5pt}$}
    \kern0.2pt}\kern1pt\vrule}\hrule}}
\newtheorem{theorem}{Theorem}[section]
\newtheorem{lemma}[theorem]{Lemma}
\newtheorem{corollary}[theorem]{Corollary}
\newtheorem{proposition}[theorem]{Proposition}
\newtheorem{remark}[theorem]{Remark}
\newtheorem{definition}[theorem]{Definition}
\newtheorem{example}[theorem]{Example}

\newcommand{\bCd}{\beq \begin{CD}}
\newcommand{\eCd}{\end{CD}\eEq}
\newcommand{\bcd}{\beq \begin{CD}}
\newcommand{\ecd}{\end{CD}\eeq}
\newcommand{\ben}{\begin{enumerate}}
\newcommand{\een}{\end{enumerate}}
\newcommand{\bEq}{\begin{eqnarray}}
\newcommand{\eEq}{\end{eqnarray}}
\newcommand{\beq}{\begin{eqnarray*}}
\newcommand{\eeq}{\end{eqnarray*}}
\newcommand{\bDf}{\begin{definition}\em}
\newcommand{\eDf}{\end{definition}}
\newcommand{\bLm}{\begin{lemma}}
\newcommand{\eLm}{\end{lemma}}
\newcommand{\bPr}{\begin{proposition}}
\newcommand{\ePr}{\end{proposition}}
\newcommand{\bTh}{\begin{theorem}}
\newcommand{\eTh}{\end{theorem}}
\newcommand{\bCr}{\begin{corollary}}
\newcommand{\eCr}{\end{corollary}}
\newcommand{\bRm}{\begin{remark}\em}
\newcommand{\eRm}{\end{remark}}
\newcommand{\bEx}{\begin{example}\em}
\newcommand{\eEx}{\end{example}}

\newcommand{\ie}{{\em i.e$.$} }
\newcommand{\eg}{{\em e.g$.$} }

\newcommand{\mto}{\mapsto}

\newcommand{\cI}{\mathcal{I}}

\newcommand{\bY}{\boldsymbol{Y}}

\newcommand{\bet}{\beta}

\newcommand{\eps}{\epsilon}

\newcommand{\tht}{\theta}

\newcommand{\Tht}{\Theta}

\title{\large{ 
{\bf Symmetry transformations of extremals and higher conserved quantities:
invariant Yang--Mills connections 
}
}
}
\author{{\normalsize Luca Accornero}
\\
{\footnotesize Department of Mathematics, University of Utrecht}
\\
{\footnotesize 3584 CD Utrecht, The Netherlands, e--mail: 
{\sc l.accornero@uu.nl
}}
\\
{\normalsize Marcella Palese\footnote{Corresponding Author}}
\\ {\footnotesize Department of Mathematics,
University of Torino}
\\
{\footnotesize via C. Alberto 10, 10123 Torino, Italy, e--mail: 
{\sc marcella.palese@unito.it}} 
}
\date{}
\begin{document}
\maketitle

\begin{abstract}  
We characterize symmetry transformations of Lagrangian extremals generating `on shell' conservation laws. We relate symmetry transformations of extremals to Jacobi fields and study symmetries of higher variations by proving that a pair given by a symmetry of the $l$-th variation of a Lagrangian and by a Jacobi field of the $s$-th variation of the same Lagrangian (with $s<l$) is associated with an `off shell' conserved current.
The conserved current associated with two symmetry transformations is constructed and, as a case of study, its expression for invariant Yang--Mills connections on Minkowski space-times is obtained.

\end{abstract}

\noindent {\bf Key words}: symmetry transformation, conservation law, second variation, Jacobi equation, Yang--Mills theory.

\noindent {\bf 2010 MSC}: 81T13,
53Z05,58A20,58E15,
58Z05.

\section{Introduction}

The description of fundamental interactions in Physics as fields associated with the action of  Lie groups on maps between manifolds has been the cornerstone of the last Century. Indeed within this picture fields are (local) maps between manifolds adapted to a fibration (which distinguishes independent from dependent variables and their peculiarity when changing coordinates), \ie are sections of fibrations, having the additional structure of a bundle (fields take values in a manifold which is the type fiber of the bundle).
In particular, it is well known that, due to their invariance properties, physical fields  can be described as sections of bundles associated with principal bundles, the configuration bundles are then the so-called gauge-natural bundle, see \eg  \cite{Ec81,KMS93}. 

It is noteworthy that the variational derivation of field equations is an intrinsic operation strictly related just to the fibration structure and its prolongation up to a given order \cite{Ec81,GoSt73,Sau89,Tra67}. 
This approach had several important developments, in particular when combined with invariance properties (geometric formulation of the Noether Theorems, specifically).

Furthermore important is now the possibility of a systematic formulation of  higher variations (see {\em Theorem \ref{th:highvar}} and {\em Theorem \ref{jacobivar}} below), which can be interpreted as variations of suitable `deformed' Lagrangians \cite{AcPa20}. 

Combined with symmetry considerations this approach extends to field theory the somehow analogous concept of so-called Sarlet--Cantrijn higher-order (dynamical) Noether symmetries in Mechanics \cite{SaCa81} (which we can think as a kind of higher order Noether--Bessel-Hagen symmetries of the Lagrangian \cite{CaPaWi16,FePaWi11}). 
The present paper refers directly to {\em canonical Noether symmetries}, \ie symmetries of the Lagrangian, which we can describe, roughly speaking, as dynamical Noether symmetries up to divergences.  Restricting to Mechanics, we recognize second order Sarlet--Cantrijn dynamical Noether symmetries up to divergences as related with second order canonical Noether currents in our approach.

Higher variations are of interest in theoretical physics, in particular concerning variations of currents \cite{FrPaWi13,Pa16}; for applications of the second variation in gravitational theory in this context see \eg \cite{FrRa02a}. 

Lagrangian symmetries and symmetries of Euler--Lagrange equations have been called by Andrzej Trautman \cite{Tra67}  invariant transformations and generalized invariant transformations, respectively, and they were characterized as particular kinds of what he called {\em symmetry transformations}, \ie transformations of extremals into extremals of the same  Euler--Lagrange equations.

Indeed, it is well known that a symmetry of Euler--Lagrange equations (generalized invariant transformations) is also a symmetry transformation of their solutions (extremals), \ie a transformation preserving the property of a field (a section of the configuration bundle on space-time) being an extremal \cite{Kru73,Tra67}.

The inverse in general is not true: symmetry transformations of solutions of equations 
could not be 
symmetries of the equations themselves. A related result stating that
a Lagrangian `dragged' along symmetry transformations of its own extremals, has the same extremals as the original Lagrangian (and an inverse statement stating that a transformation dragging a given Lagrangian in a Lagrangian having the same extremals is a symmetry transformation of the extremals) was obtained in \cite{Kru15b} (see Theorem  \ref{Krupka} below).

We focus on conservation laws by explicating the relation among higher variations of Lagrangians, symmetry transformations of extremals, Jacobi fields, and conserved currents. 
We characterize symmetry transformations of extremals as particular transformations of the Euler--Lagrange forms to source forms vanishing along extremals of the original Lagrangian and specifically as Jacobi fields {\em along extremals} ({\em Theorem \ref{MAIN}}). 

Compared with generalized symmetry transformations (\ie transformations leaving invariant the Euler--Lagrange form of a Lagrangian) such transformations provide a weaker  invariance property, since the Euler--Lagrange form is not invariant under their action, although it is transformed to a source form having the same extremals.  Therefore the equations change, but the solutions of the one equation are also solutions of the second {\em and vice versa}.

By explicating the relation among the variation of an Euler-Lagrange form with the second variation of a Lagrangian and with the Jacobi morphisms ({\em Proposition \ref{th:conslaws} } and {\em Remark \ref{REMARK}}), we prove that with this sort of weaker invariance is anyway associated a conserved current, and in particular that this current can be identified as a Noether current for a Lagrangian `deformed'  by a symmetry transformation of extremals  and associated with (or generated by) a symmetry transformation of extremals.
More specifically, symmetry transformations of extremals generate conserved currents along the extremals themselves.
Indeed {\em Theorem \ref{Jacobicons} } states the existence of a weak (\ie along extremals) conservation law for any couple of (infinitesimal) generators of (vertical) symmetry transformations. 

As an explicit example, we write the expressions of the on shell conserved current generated by couples of symmetry transformations of extremals.

\section{Contact structure, geometric integration by parts and the `representation' problem}\label{section2}

Throughout this paper, we work with a fibration $\pi:Y\to X$, where $Y$ is an $m+n$ dimensional manifold and $X$ is an $n$-dimensional manifold. When choosing coordinates, we will always pick fibered coordinates $(x^i, y^\sigma)$ defined over open subsets $\pi^{-1}(U)\subset Y$, where $U$ is an open subset of the base. We also set $ds$ to be the local expression of a volume element $dx^1\wedge\ldots\wedge dx^n$ on $X$ and $ds_i=\frac{\partial}{\partial x^i}\rfloor ds$. Given this geometric setting, physical fields are encoded as local sections of the surjective submersion $\pi:Y\to X$. 

Let $\Omega_q(J^kY)$ denote the module of $q$-forms on the set $J^kY$ of equivalence classes of (local) sections of the fibration having a contact of order $k$ in a point. Note that  $J^kY$ as the structure of a differentiable manifold and the structure of a fibration $\pi_{k }: J^kY \to X$, the prolongation of order $k$ of $\pi :Y\to X$.

A prolongation map $j^k$ assigns to (local) sections of the fibration $\pi: Y\to X$ (local) sections of the fibration $\pi_k: J^kY\to X$. If $\gamma$ is a section of $\pi$, $j^k\gamma$ is defined as the map assigning to $x\in X$ the $k$-jet $j^k_x\gamma$ of $\sigma$ at $x$.
A differential $q$-form $\alpha \in \Omega_q(J^kY)$ is called contact if $j^k\gamma^*(\alpha)=0$ for all sections $\gamma$ of $\pi$. It is easy to see that forms $\omega$  locally given as
\beq 
\omega^\sigma_{j_1\dots j_h}= dy^\sigma_{j_1\dots j_h}-y^\sigma_{j_1\dots j_h i}dx^i \,
\eeq 
for $0\leq h <k$ are indeed  contact $1$-forms.  

In particular $(dx^i,\omega^\sigma,\omega^\sigma_{j_1},\dots, \omega^\sigma_{j_1\dots j_{k-1}}, dy^\sigma_{j_1 \dots j_k})$ is a local basis for $1$-forms on $J^kY$. Contact forms on a fixed prolongation space $J^kY$ generate an ideal of the exterior algebra; this is usually called the contact structure induced by the affine bundle projections $\pi_{k,k-1}: J^kY\to J^{k-1}Y$, see \eg \cite{Sau89,Kru73}. It is important to notice that if $\alpha$ is contact so is $d\alpha$; on the other hand, the ideal of the exterior algebra of forms on $J^kY$ generated by the forms $\omega^\sigma_{j_1},\dots, \omega^\sigma_{j_1\dots j_{k-1}}$ is not closed under exterior derivation. 

For every form $\rho \in \Omega_q(J^kY)$, by the contact structure we obtain the canonical decomposition  \cite{Kru73}
\beq \pi_{k+1,k}^*(\rho)=p_0\rho+p_1\rho+\dots+p_q\rho
\eeq 
where $p_0\rho$ is a form that is horizontal on $X$ (and so is often denoted by $h\rho$) while $p_i\rho$ is an $i$-contact $q$-form, that is a form generated by wedge products containing exactly $i$ contact $1$-forms.  We remark that if $q>n$ every $q$-form $\rho$ is contact; then we call it {\em strongly} contact if $p_{q-n}=0$. 
The contact structure induces also the splitting of the exterior differential $\pi_{k+1,k}^* d \rho= d_H\rho+ d_V\rho$ in the so called horizontal and vertical differentials, given by  $d_H\rho=\sum\limits^{q}_{l=0}p_ldp_l \rho$ and $d_V\rho=\sum\limits^{q}_{l=0}p_{l+1}dp_l \rho$, respectively. 

According to \cite{KrMu05}  we define the {\em formal derivative} with respect to the $i$-th coordinate, $i=1, \dots , n$, by an abuse of notation also denoted by $d_i$, as  an operator acting on forms. Explicitly, we require $d_i$ to be the usual total derivative on $0$-forms, to commute with the exterior derivative and to satisfy the Leibniz rule with respect to the wedge product. We see that $d_H\rho=(-1)^qd_i\rho \wedge dx^i$ if $\rho$ is a $q$-form. On the  basis $1$-forms associated to the contact structure we have
$d_idx^j=0$, $d_i \omega^\sigma_{j_1\dots j_r}=\omega^\sigma_{j_1\dots j_r i}$, $d_idy^\sigma= dy^\sigma_i$. Notice that $d_i$ induces a vector field on $J^kY$ along $\pi_{k+1,k}$; we still denote it by $d_i$ and refer to it as the formal derivative.

In the following a multi-index will be an ordered $s$-uple $I=(i_1, \dots i_s)$; the length of $I$ is given by the number $s$;
and an expression such as  $Ij$ denotes the multi-index given by the $(s+1)$-uple $(i_1, \dots i_s, j )$.

As much as  the integration by parts procedure is concerned, we will use the local formula 
$\omega^\sigma_{Ii}\wedge ds=-d\omega_I\wedge ds_i$. We also recall the properties $d_J\omega^\sigma=\omega^\sigma_J$ and $\frac{\partial}{\partial y^\nu_J}\rfloor \omega^\sigma_I=\delta^\sigma_\nu \delta^J_I$ (where the Kronecker symbol with multi-indices has the obvious meaning: it is $1$ if the multi-indices coincide up to a rearrangement and $0$ otherwise). 

Finally, if $\psi$ is a projectable vector field on $Y$ (\ie an infinitesimal automorphism preserving the fibration), $j^k\psi$ is the projectable vector field, defined on  $J^kY$, associated with the prolongation of the flow of $\psi$ (see, \eg  \cite{Kru73,Sau89,Tra67}).


The contact structure of jet prolongations enables to define an algebro-geometric object deeply related to the calculus of variations: a differential sequence of sheaves made of equivalence classes of differential forms taking a variational meaning. We refer to \cite{Kru90}, where the construction of a sequence of `variational sheaves' can be found and to \cite{KrMu05,PaRoWiMu16} for the representation of finite order variational sequences.
 The concept of a sheaf is due to Leray;
 a classical reference on this topic is \eg \cite{Bre67}.

Let $\Omega^k_q$ denote the sheaf of differential $q$-forms on $J^kY$. It can be seen as a  sheaf on $Y$; in fact we assign to an open set $W\subseteq Y$ a form defined on $\pi^{-1}_{r,0}(W)$. We set $\Omega^k_{0,c}=\{0\}$ and denote by $\Omega^k_{q,c}$ the sheaf of contact $q$ forms, for $q\leq n$, or the sheaf of strongly contact q-forms if $q>n$. 
The quotient sequence of the de Rham sequence of forms
\beq 
\{0\} \to \mathbb{R}_Y\to \Omega^k_0\to  \dots \to \Omega^k_{n}/\Theta^k_{n}\to\Omega^k_{n+1}/\Theta^k_{n+1}\to
 \dots \to \Omega^k_N\to \{0\}\,,
\eeq 
where $\Theta^k_q=\Omega^k_{q,c}+d\Omega^k_{q-1,c}$, $N=\dim(J^kY)$ and $\mathbb{R}_Y$ is the constant sheaf over $Y$,
 is called  the {\em Krupka's variational sequence of order $k$} \cite{Kru90}. Let us denote the quotient sheaves by $\mathcal{V}^k_q$. Morphisms in this sequence  are denoted by $\mathcal{E}_q: \mathcal{V}^k_q \to \mathcal{V}^k_{q+1}$ and they
  are quotients of the exterior differential  \ie $\mathcal{E}_q([\rho])=[d\rho]$.
By this construction classes of forms modulo contact forms 
 are interpreted as differential forms relevant for calculus of variation (Lagrangians, currents, source forms and so on); the Euler--Lagrange mapping can be identified with a morphism in the variational sequence. The representation of the second variational derivative has been studied from this point of view 
 \cite{FFPW11,FPV05,PaWi03,PaWi07}.

The interest of this construction in Physics is that the cohomology of the complex of global section of the variational sequence is the de Rham cohomology of $Y$ \cite{Kru90}. 
Dealing with exact sequence of sheaves and resolutions enables to study cohomology  obstructions to {\em variational exactness of variationally closed forms} and this turns out to be of interest in many different areas of Physics; for example an obstruction to the existence of global extremals is related to the obstruction to the existence of global Noether--Bessel-Hagen currents \cite{PaWi17}. 

Strictly related to concrete applications is then  the so called {\em representation problem}, which, roughly speaking, consists in showing that {\em classes of forms}, \ie elements of the quotient groups $\mathcal{V}^r_q$, can be associated with {\em global differential forms}. 
By the intrinsic geometric structure of the calculus of variations on finite order prolongations of fibrations, indeed, it is possible to define an operator (called {\em representation mapping}) which takes a differential forms on the prolongation of order $r$ and associate to it a differential form on a certain prolongation order $s\geq r$, having a meaning in the Lagrangian formalism for field theory, \ie 
$R^r_q: \Omega^r_q \to \Psi^s_q$, with $\Psi^s_q$ an abelian group of forms of order $s$, 
such that $\ker R^r_q=\Theta^r_q$. It provides
an isomorphism $\mathcal{V}^r_q \cong \Psi^s_q=R^r_q(\Omega^r_q)$.

For $q\leq n$, $R^r_q$ can be taken to be simply the `horizontalization' $h=p_0$. For $q\geq n+1$, it is the image of an operator denoted by $\mathcal{I}$, which will be suitably defined below and which reflects in an intrinsic way the procedure of getting a distinguished representative of a class $[\rho] \in \Omega^k_{q}/\Theta^k_{q}$ for $q > n$ by applying to $\rho$ the operator $p_{q-n}$ and then factorizing by 
$\Theta^k_{q}$, see \eg \cite{PaRoWiMu16}.

In this paper we will refer to the {\em interior Euler operator} defined within the finite order variational sequence according to \cite{KrMu03,KrMu05} and applied to the representation of variational Lie derivatives according to \cite{PaRoWiMu16}.  
\bDf
In the following, differential forms which are $\omega^\sigma$ generated $l$-contact $(n+l)$-forms will be called {\em  source forms}. 
\eDf
Now define locally the map
$\mathcal{I}:\Omega^r_{n+k} \to \Omega^{2r+1}_{n+k} $
by 
\beq
\mathcal{I}(\rho)=\frac{1}{k}\omega^\sigma \wedge I_\sigma=\frac{1}{k}\omega^\sigma\wedge\sum^{r}_{|I|=0}(-1)^{|I|}d_I(\frac{\partial}{\partial y^\sigma_I}\rfloor p_k\rho) \,.
\eeq
For a given $\rho$, $\mathcal{I}(\rho)$ is a source form of degree $n+k$ and  it is by construction a $k$-contact form. It turns out that, if $\rho$ is global, $\mathcal{I}(\rho)$ is a globally defined form; for a proof, see \cite{KrMu05}.

In view of a characterization of Noether currents, we study the difference between $(\pi_{2r+1,r+1})^*(p_k\rho)$ and $\mathcal{I}(\rho)$.
In particular, we define the {\em residual operator} $\mathcal{R}$
by the following decomposition formula which is in fact a {\em geometric integration parts formula} 
\bEq \label{integration}
(\pi_{2r+1,r+1})^*(p_k\rho)=\mathcal{I}(\rho) + p_kdp_k\mathcal{R}(\rho) \,.
\eEq 
Note that, although the decomposition above has a global meaning, $\mathcal{R}(\rho)$ is a strongly contact ($n+k-1$)-form defined only locally. 

\bEx \label{residual_operator}

Following \cite{KrMu05} we can characterize $\mathcal{R}(\rho)$ in local coordinates. For $k\geq1$, if $\Psi^I_\sigma$ are $(k-1)$-contact $(n+k-1)$-forms and  if $\omega^\sigma_I$ are local generators of contact $1$-forms, up to pull-backs, we can write (a sort of integration by parts on formal derivatives of forms)
\beq
 p_k\rho=\sum^{r}\limits_{|I|=0} \omega^\sigma_I \wedge \Psi^I_\sigma  =\sum\limits^r_{|I|=0}d_I(\omega^\sigma\wedge \zeta^I_\sigma) = \mathcal{I}(\rho) + p_kdp_k\mathcal{R}(\rho) \,,
\eeq 
with $\zeta^I_\sigma=\sum\limits^{r-|I|}_{|J|=0}(-1)^{|J|}{{|I|+|J|}\choose{|J|}}d_J\Psi^{JI}_\sigma  \,.
$
The first term gives us the Euler--Lagrange form, while
by rewriting $\omega^\sigma\wedge \zeta^I_\sigma=\Phi^I\wedge ds$, for suitable $k$-contact $k$-forms $\Phi^I$ on $J^{2r}Y$, we get 
\beq
\sum\limits^r_{|I|=1}d_I(\omega^\sigma\wedge \zeta^I_\sigma)= d_{H} (\sum\limits^{r-1}_{|I|=0}(-1)^k d_I\Phi^{Ij}\wedge ds_j )= d_{H}\mathcal{R}(\rho) \,.
\eeq
\eEx
This local expressions for $\mathcal{R}(\rho)$ will be exploited in Example \ref{Jac_conservation_ex} for the case $k=1$, specifically for concrete  $1$-contact $(n+1)$-forms $\omega^\sigma_I \wedge \Psi^I_\sigma$ associated with the exterior differential of a suitably `deformed' Yang--Mills Lagrangian. We will write explicitly the forms $\Phi^{Ij}$ relative to this Lagrangian. Combined with results of Theorem \ref{Jacobicons}, this approach will enable us to obtain explicit conserved currents associated with symmetry transformations of Yang--Mills extremals on Minkowski space-times.

\section{Higher variations and related currents} \label{highervariations}

The representation by the horizontalization $h$ and by the interior Euler operator $\cI$ (also called the
Takens 
representation \cite{PaRoWiMu16}) defines a sequence of sheaves of {\em differential forms} (rather than of classes of differential forms), such that both the objects and the morphisms have a straightforward interpretation in the calculus of variations. We can obtain formulae for (higher) variations of a Lagrangian based on an iteration of the first variation formula expressed through this representation.

\subsection{Noether currents}\label{3.1}

The formulation of the First Noether Theorem \cite{Noe18} is concerned with 
the representation of {\em variational Lie derivatives} of classes of degree $n$, which  illustrates the relation between the interior Euler operator, the Euler--Lagrange operator and the exterior differential, as well as the emerging of the divergence of the Noether currents by contact decompositions and geometric integration by parts formulae.

In the following, for any $n$-form $\rho$, $\cI (d \rho) =\mathcal{I}(dh\rho) = \cI d(h\rho)$ is the 
 Euler--Lagrange form  $E_n  (h\rho)$, while
 $hdh\mu=p_0 dp_0\mu$ is the horizontal differential $d_H(h\mu)$, which can be recognized as a divergence
(for the notation and the interpretation in the context of geometric calculus of variations more details can be found  \eg  in \cite{PaRoWiMu16}).
\bTh\label{var_lie_der} 
For  any $n$-form $\rho$ and for any $\pi$-projectable vector field $\psi$ on $\bY$, we have, up to pull-backs by projections,
\bEq\label{eq:var_n}
L_{J^{r+1}\psi}h\rho=\psi_V\rfloor \cI d(h\rho)+ 
d_H(J^{r+1}\psi_V\rfloor p_{d_Vh\rho} + \psi_H\rfloor h\rho)  \,
\eEq
where
$ p_{d_Vh\rho} = - p_1\mathcal{R}(dh\rho)$.
\eTh

The formula above has been first obtained by Noether in the proof of her celebrated First Theorem (see the original Noether paper in the historical survey \cite{Kos11}).
This suggest the definition of a Noether current.
\bDf
The {\em Noether current for a Lagrangian $\lambda$ associated with $\psi$} is defined as 
\beq
\epsilon_{\psi}(\lambda)=J^{r+1}\psi_V\rfloor p_{d_V\lambda} + \psi_H\rfloor \lambda\,.
\eeq
The term $p_{d_V\lambda} = - p_1\mathcal{R}(d\lambda)$ is called {\em a local generalized momentum}.
\eDf
It should be stressed that a Noether current is defined for any projectable vector field, independently from it being a Lagrangian symmetry or not. When it is not a symmetry of course the Noether current is {\em not} conserved along critical sections. 

A generalization of formula \eqref{eq:var_n} to class of degree greater or lower than $n$ has been obtained \cite{CaPaWi16,PaRoWiMu16,PaRoZa20}.

\subsection{Higher Noether currents}

We now tackle a systematic formulation of  higher variations, interpretable as variations of suitable `deformed' Lagrangians \cite{AcPa20}; combined with symmetry considerations this approach extends to field theory the concept of so-called higher-order Noether symmetries in Mechanics developed in \cite{SaCa81}.

Now we obtain a formula for the second variation, which will be further exploited in section \ref{Symmetrysection}. 
We note that $L_{J^{r+1}\psi}h\rho=hL_{J^r\psi}\rho$, and then apply a standard inductive reasoning. Of course, the iterated variation is pulled-back up to a suitable order, in order to suitably split the Lie derivatives \cite{AcPa20}. 

\bTh\label{2nd_var}
For  any $n$-form $\rho$ and any pair of  $\pi$-projectable vector fields $\psi_1$ and $\psi_2$, we have, up to pull-backs by projections,
\bEq \label{eq:2nd_var}
L_{J^{r+1}\psi_2} L_{J^{r+1}\psi_1}h\rho =\psi_{2,V}\rfloor \cI d(\psi_{1,V}\rfloor \cI d(h\rho)) + \\
+d_H\epsilon_{\psi_2}(\psi_{1,V}\rfloor \cI d(h\rho))+d_H\epsilon_{\psi_2}(d_H\epsilon_{\psi_1}(h\rho)) \nonumber
\eEq
where we define the following (higher) Noether currents associated with $\psi_2$ for the respective new Lagrangians:
\bEq\label{higher noether current1}
 \epsilon_{\psi_2}(\psi_{1,V}\rfloor \cI d(h\rho))=\psi_{2,H}\rfloor \psi_{1,V}\rfloor \cI d(h\rho) +
J^{r+1}\psi_{2,V}\rfloor p_{d_V \psi_{1,V}\rfloor \cI d(h\rho)}  \,, \\ 
 \epsilon_{\psi_2}(d_H\epsilon_{\psi_1}(h\rho))=\psi_{2,H}\rfloor d_H(J^{r+1}\psi_{1,V}\rfloor p_{d_Vh\rho}+\psi_{1,H}\rfloor h\rho)+ \label{higher noether current2}\\ 
+J^{r+1}\psi_{2,V}\rfloor p_{d_V d_H (J^{r+1}\psi_{1,V}\rfloor p_{d_V h\rho}+\psi_{1,H}\rfloor h\rho)} \,. \nonumber
\eEq
\eTh
Note that the expression \eqref{eq:2nd_var} is given in terms of $\mathcal{I}$ and $\mathcal{R}$.  

Related to this  formula is an identity which will suggest the definition of the  Jacobi morphism, with a look to a specific characterization of symmetry transformations of extremals (see Definition \ref{Jac}).

Let  then $\psi_1$, $\psi_2$ be {\em vertical vector fields}. We note that, due to the exactness of  the representation sequence and by linearity of the Lie derivative, (for $s$ a suitable prolongation order) we can write
\bEq \label{LieDer}
J^s\psi_{1}\rfloor L_{J^{s}\psi_2}\cI d(h\rho) = \psi_{1}\rfloor \cI d (\psi_2 \rfloor \cI d(h\rho)) = \\ = L_{J^{s}\psi_2} L_{J^{s}\psi_1}h\rho - [\psi_2,\psi_{1}]\rfloor \cI d(h\rho)-d_H\epsilon_{\psi_2}(d_H\epsilon_{\psi_1}(h\rho)) \,. \nonumber
\eEq 
 From \eqref{eq:2nd_var} we get then  the following identity.

\bPr\label{th:conslaws}
For every pair of vertical vector fields $\psi_1$ and $\psi_2$ it holds
\bEq\label{eq:2nd_commu}
\psi_{1}\rfloor \cI d (\psi_{2}\rfloor \cI d(h\rho)) -\psi_{2}\rfloor \cI d(\psi_{1}\rfloor \cI d(h\rho)) = \\ 
= [\psi_1,\psi_{2}]\rfloor \cI d(h\rho) + d_H(\epsilon_{\psi_2}(\psi_{1}\rfloor \cI d(h\rho))) \,.\nonumber
\eEq
\ePr
Note that, being the vector fields vertical, here we have $\epsilon_{\psi_2}(\psi_{1}\rfloor \cI d(h\rho))=
J^{r+1}\psi_{2}\rfloor p_{d_V \psi_{1}\rfloor \cI d(h\rho)} $. Note also that this current is the Noether current for the `deformed' Lagrangian $\psi_{1}\rfloor \cI d(h\rho)$ and associated to $\psi_2$.

\bRm
As we already mentioned in the Introduction there exists a concept of a higher order Noether symmetry referring actually to a higher order generalization of a `Noether symmetry' intended as a symmetry of the exterior differential of the Poincar\'e--Cartan equivalent of a given Lagrangian, \ie $\Psi$ is called a `Noether symmetry' if $L_\Psi d\tht=0$. In particular, we refer to the generalization due to
Sarlet--Cantrijn \cite{SaCa81}, and in the following we shall  call  a `Noether symmetry' according to them as a Sarlet--Cantrijn symmetry.

 We stress that this symmetry differs from a Noether symmetry as a symmetry of the Lagrangian (due to the fact that the Poincar\'e--Cartan equivalent of $L$ differs for a $1$-contact term, let us call it $\Omega$), which is the original meaning used also by Emmy Noether who referred to symmetries of the action, \ie of the Lagrangian.  As well known symmetries of the Lagrangian are {\em also} symmetries of the Euler--Lagrange form (which can be  expressed, in terms of the differential of the Poincar\'e--Cartan equivalent as as $p_1d\tht$), but the converse is not true in general.  Indeed symmetries of the Euler--Lagrange form are generalized symmetries of the Lagrangian, \ie symmetries up to a horizontal differential. Therefore, Sarlet--Cantrijn symmetries, which are symmetries of the Poincar\'e--Cartan equivalent  up to a differential can be identified as a kind of generalized symmetries. 
 Accordingly higher Noether symmetries and currents in this paper can be compared with Sarlet--Cantrijn ones.
 \eRm

\bEx
Let us now denote by  $\tht =  L +\Omega$ the Poincar\'e--Cartan equivalent 
of the Lagrangian $L=h\rho$. Let us for a moment skip the prolongation symbols to simplify the notation; we have the following.

Let $\Psi$ denote a projectable vector field and let  $L_\Psi L_\Psi d\tht = d(\Psi\rfloor d(\Psi\rfloor d\tht))= 0$.
On the one hand, by the naturality of the Lie derivative and by Equation \eqref{LieDer}  we have 
\beq
& & L_\Psi L_\Psi d\tht = d L_\Psi L_\Psi \tht = \\
& & = d [\Psi_V \rfloor \cI d (\Psi_V \rfloor \cI d L) + d_H \eps_\Psi (\Psi_V \rfloor \cI d L) + d_{H} \eps_\Psi (d_H\eps_\Psi(L)) +L_\Psi L_\Psi \Omega]\,.
\eeq
Let us take  the quotient  modulo the contact structure $\Tht$, $\cI d L_\Psi L_\Psi \tht $. Since $ \cI d L_\Psi L_\Psi \Omega = L_\Psi L_\Psi \cI d\Omega=0$, being $\Omega $ a contact form and being $d\Omega$ also in the contact sheaf $\Theta$.
Thus we have  $\cI d L_\Psi L_\Psi \tht = \cI d \eta $, with $\eta = \Psi_V \rfloor \cI d (\Psi_V \rfloor \cI d L) + d_H \eps_\Psi (\Psi_V \rfloor \cI d L) +  d_H\eps_\Psi (d_H\eps_\Psi(L)) $.

Note that under the assumption that $L_\Psi L_\Psi d\tht = 0$,
 $\eta$ is  a $\cI d $-closed form, \ie  $\cI d \eta = 0$, then by the exactness of the variational sequence, we have locally  $\Psi_V \rfloor \cI d (\Psi_V \rfloor \cI d L) = d_H  ( G- [\eps_\Psi (\Psi_V \rfloor \cI d L) +  \eps_\Psi (d_H\eps_\Psi(L))])$.

On the other  hand, following \cite{SaCa81}, let $\bet = \Psi\rfloor d\alpha$, where $\alpha=\Psi\rfloor d \theta$. We have $L_\Psi L_\Psi \tht = \bet +d (\Psi\rfloor d (\Psi\rfloor \theta))=\bet +d \xi$; and from the above we also get $\cI d \eta = \cI d( \bet +d \xi)=  \cI d ( \bet + d _V \xi)$.

We now elaborate and compare these two issues. 
Indeed, from Proposition \ref{th:conslaws}  we have the identity $d_H \eps_{\Psi_V} (\Psi_V \rfloor \cI d L)=0$. Thus 
\beq
\Psi_V \rfloor \cI d (\Psi_V \rfloor \cI d L) = d_H  ( G - \Psi_H \rfloor \Psi_V \rfloor \cI d L -  \eps_\Psi (d_H\eps_\Psi(L))]) \,.
\eeq
Furthermore, if we assume $\Psi_V$ to be such that 
$\Psi_V \rfloor \cI d (\Psi_V \rfloor \cI d L) =0$ (\ie to be a Jacobi field, see later) then we get the conservation law 
\beq
d_H (G  -\Psi_H \rfloor \Psi_V \rfloor \cI d L -\eps_\Psi (d_H\eps_\Psi(L))) =0 \,,
\eeq
which along an extremal reads
\beq
d_H (G  -\eps_\Psi (d_H\eps_\Psi(L))) =0 \,.
\eeq

On the other hand since $ L_\Psi L_\Psi d\tht = d(\Psi\rfloor d (\Psi\rfloor d\tht) )=0$ then $d  \bet =0$, therefore locally $\bet =d F$ \cite{SaCa81}.

By taking the horizontal part we have $h\bet =d_H F$. But  from the above $d G=\eta=\bet +d\xi$ thus we can take $G= F+\xi$, and we have locally that, for $\Psi_V$ a Jacobi field along an extremal
\beq
d_H F= d_H( \eps_\Psi (d_H\eps_\Psi(L) )- \xi) =   d_H( \eps_\Psi (d_H\eps_\Psi(L)) - \Psi\rfloor d (\Psi\rfloor \tht)  \,.
\eeq
Now, up to pull-backs and jet prolongations of the vector field $\Psi$, we have
\beq
& & d_H F= 
 d_H(  \Psi_{H}\rfloor d_H( \Psi_{V}\rfloor p_{d_V L}+\Psi_{H}\rfloor L )+\Psi_{V}\rfloor p_{d_V d_H (\Psi_{V}\rfloor p_{d_V L}+\Psi_{H}\rfloor L)} + \\ 
& & -\Psi_H \rfloor (d_H(\Psi_{V}\rfloor p_{d_V L}+\Psi_{H}\rfloor L) ) -\Psi_H \rfloor (d_V(\Psi_{V}\rfloor p_{d_V L}+\Psi_{H}\rfloor L)) + \\
& & - \Psi_V \rfloor (d_H(\Psi_{V}\rfloor p_{d_V L}+\Psi_{H}\rfloor L) ) - \Psi_V \rfloor (d_V(\Psi_{V}\rfloor p_{d_V L}+\Psi_{H}\rfloor L) ) )  =   \\
& &d_H( \Psi_{V}\rfloor p_{d_V d_H (\Psi_{V}\rfloor p_{d_V L}+\Psi_{H}\rfloor L)} -\Psi_H \rfloor (d_V(\Psi_{V}\rfloor p_{d_V L}+\Psi_{H}\rfloor L)) + \\
& & - \Psi_V \rfloor (d_H(\Psi_{V}\rfloor p_{d_V L}+\Psi_{H}\rfloor L) ) - \Psi_V \rfloor (d_V(\Psi_{V}\rfloor p_{d_V L}+\Psi_{H}\rfloor L) ) ) =\\
& & =   d_H(  \Psi_{V}\rfloor p_{d_V d_H (\Psi_{V}\rfloor p_{d_V L}+\Psi_{H}\rfloor L)} 
 - \Psi_V \rfloor (d_V(\Psi_{V}\rfloor p_{d_V L}+\Psi_{H}\rfloor L) )\\
 & & =   d_H(    \eps_{\Psi_V} (d_H\eps_{\Psi_V}(L)+   \Psi_{V}\rfloor p_{d_V d_H (\Psi_{H}\rfloor L)} 
 - \Psi_V \rfloor (d_V(\Psi_{V}\rfloor p_{d_V L}+\Psi_{H}\rfloor L) ) \,.
 \eeq
which explicates the relationship between Sarlet--Cantrijn second-order Noether conserved current and the conserved current along extremals $\eps_{\Psi_V} (d_H\eps_{\Psi_V}(L))$, for $\Psi_V$ a Jacobi field; see later Equation \eqref{eq:Jacobicons1} combined with Proposition \ref{th:conslaws}. The last term can be shown to vanish under the horizontal differential \cite{CaPaWi16,PaRoWiMu16,PaRoZa20}, thus in the case of a {\em vertical} Sarlet--Cantrijn symmetry the conserved current $F$ essentially coincides with the Noether conserved current $\eps_{\Psi_V} (L_{\Psi_V}(h\rho))$ up to a horizontal differential.
\eEx

By Theorems \ref{var_lie_der} and  \ref{2nd_var} formulae for the higher variations of $h\rho$ are obtained \cite{Ac17}.
\bTh \label{th:highvar}
Let $\rho$ be an $n$-form on $J^kY$. Consider the Lagrangian $h\rho$ and take $l$ variation vector fields $\psi_1, \dots, \psi_l$. Define recursively a sequence $r_l$ by 
\beq r_l=2r_{l-1}+1,\ \ \ r_0=r
\eeq 
We have
\begin{equation}\label{eq:highvar}
\begin{aligned}
(\pi_{r_l,r+1})^*\left(L_{J^{r+1}\psi_l}\right.&\left.\dots L_{J^{r+1}\psi_1}h\rho\right)=\\
=&\psi_{l,V}\rfloor \cI d(\psi_{l-1,V}\rfloor \cI d(\dots \psi_{2,V} \rfloor \cI d(\psi_{1,V}\rfloor \cI d(h\rho))\dots)+\\
+&d_H\epsilon_{\psi_l}(\psi_{l-1,V}\rfloor \cI d(\dots \psi_{2,V} \rfloor \cI d(\psi_{1,V}\rfloor \cI d(h\rho))\dots)+\\
+&d_H\epsilon_{\psi_l}(d_H\epsilon_{\psi_{l-1}}(\psi_{l-2,V}\rfloor \cI d(\dots (\psi_{1,V}\rfloor \cI d(h\rho))\dots)+\\
&\dots\\
+&d_H\epsilon_{\psi_l}(d_H\epsilon_{\psi_{l-1}}d_H(\dots d_H\epsilon_{\psi_2}(d_H\epsilon_{\psi_1}(h\rho))\dots) \,.
\end{aligned}
\end{equation}
\eTh
\bPf
The proof is a straightforward induction using as base step the case $l=1$ 
or $l=2$. Taking into account the exactness of the representation sequence, the inductive step follows easily.
\ePf

\bRm

By means of a recursive application of \eqref{eq:var_n} or \eqref{eq:2nd_var}, the currents  which appear in formula \eqref{eq:highvar} can be worked out more explicitly and characterized as Noether currents similarly to the expressions in   \eqref{higher noether current1} and  \eqref{higher noether current2}.
\eRm

The variation of any order of a Lagrangian $h\rho$ is a horizontal form, \ie again a Lagrangian,  therefore
we can express its variation by means of formula \eqref{eq:var_n}. On the other hand formula \eqref{eq:highvar} gives us the possibility to investigate how to relate the {\em symmetries} of a variation of $h\rho$ to $h\rho$ itself (see Theorem \ref{jacobivar} below).

\section{Symmetry transformations of extremals and conserved currents}\label{Symmetrysection}

Let $h\rho$ be a Lagrangian of order $r+1$ on $Y$.
\bDf
A (local) section $\gamma$ is an extremal of $h\rho$ if it satisfies 
\beq
\cI d(h\rho) \circ J^{2r+1}\gamma = 0 \,.
\eeq
\eDf

Let now $\phi$ be an automorphism of $Y$ (\ie a transformation preserving the fibration) with projection $\phi_0$, and let $J^{r+1}\phi$ be its prolongation.

\bDf
The automorphism $\phi$ is a symmetry transformation of an extremal $\gamma$, if the section $\phi\circ\gamma\circ \phi_{0}^{-1}$ is also an extremal. \ie
\beq
\cI d(h\rho) \circ J^{2r+1}(\phi\circ\gamma\circ \phi_{0}^{-1}) = 0 \,.
\eeq
\eDf

A $\pi$-projectable vector field $\psi$ is the generator of symmetry transformations of $\gamma$, if its local one-parameter group of transformations is a flow of symmetry transformations of $\gamma$.
It can be shown that a symmetry of $\cI d(h\rho) $ is also a symmetry transformation of every extremal  $\gamma$ \cite{Tra67,Kru73}.

According with the above references, the following relates symmetry transformations of extremals with projectable vector fields dragging $h\rho$ in such a way that $L_{J^{r+1}\psi}h\rho$ admits the same extremals.
\bTh\label{Krupka}
Let $h\rho$ be a Lagrangian of order $r+1$ and let $\gamma$ be an extremal. Then a $\pi$-projectable vector field $\psi$ generates symmetry transformations of $\gamma$ if and only if 
\beq
\cI d(L_{J^{r+1}\psi}h\rho) \circ J^{2r+1}\gamma = 0 \,,
\eeq
\eTh

\bRm\label{IMPORTANT REMARK}
We make now an observation which will have a fundamental consequence when related to Theorem \ref{2nd_var}. Indeed, we note that, being the Lie derivative a natural operator, it holds $L_{J^{2r+1}\psi} \cI d(h\rho) = \cI d(L_{J^{r+1}\psi}h\rho)$ and $\psi$ generates symmetry transformations of $\gamma$ if and only if 
\beq
(L_{J^{2r+1}\psi} \cI d(h\rho) ) \circ J^{2r+1}\gamma =  0 \,.
\eeq
\eRm
We thus characterize vertical symmetry transformations of extremals as particular transformations of the Euler--Lagrange forms to source forms vanishing along extremals of the original Lagrangian. 

We are now able to state our {\em first main result}, which is the premise for the next fundamental step: to characterize vertical symmetry transformations of extremals specifically as Jacobi fields along extremals (see Theorem \ref{th:Jacobi}  and Remark \ref{REMARK} below). 
We do this basically by expressing the Lie derivative of Euler--Lagrange forms in terms of the second variation (see also \cite{PaWi03}).

\bTh \label{MAIN}
Let $h\rho$ be a Lagrangian of order $r+1$, let $\gamma$ be an extremal. Then a {\em vertical} vector field $\psi$ generates vertical symmetry transformations of $\gamma$ if and only if  
\beq
\cI d (\psi \rfloor \cI d(h\rho))  \circ J^{4r+1}\gamma = 0 \,.
\eeq
\eTh
\bPf
The result follows from Theorem \ref{Krupka} and Remark \ref{IMPORTANT REMARK} by using
the identity \eqref{LieDer} which, we stress, holds true for any vertical vector field $\psi_1$.
\ePf

We focus on higher order Noether currents and in particular on currents associated with the infinitesimal second variation formula \eqref{eq:2nd_var} in a specific way.

Roughly speaking, {\em up to horizontal differentials}, the second variation (generated by vertical vector fields) of a Lagrangian $\lambda$ is the Jacobi morphism (see \cite{GoSt73} for first order field theory; see also \cite{FPV05}). Here we define the Jacobi morphism within the representation
sequence, \ie by the interior Euler operator.  
\bDf \label{Jac}
Let  $X_V(Y)$ be the space of vertical vector fields on $Y$. 
The map 
\bEq\label{eq:Jacobi}
& &\mathcal{J}: \Omega^r_{n,X}(J^rY) \to X^*_V(J^{2r+1}Y)\otimes X^*_V(Y)\otimes \Omega^r_{n,X}(J^rY)\\
& & \lambda : \, \mto \bullet\ \rfloor \cI d( \bullet\ \rfloor \cI d(\lambda)) 
\eEq
is called the Jacobi morphism associated with the Lagrangian $\lambda$ .
\eDf
The Jacobi morphism is self-adjoint along critical sections of a Lagrangian field theory of any order (see also \cite{AcPa20,FPV05}). 
This is a property of great importance in physical applications.

\bTh\label{th:Jacobi}
For any pair  of vertical  vector fields $\psi_1$, $\psi_2$ on $Y$, we have
\beq 
J^{2r+1}\psi_2\rfloor\mathcal{I}(J^{2r+1}\psi_1\rfloor d \mathcal{I}(d\lambda))=0\,.
\eeq 
Along extremals the Jacobi morphism is self adjoint. 
\eTh
Indeed we have
\bEq\label{eq:coordJacobi2}
\cI d(\psi\rfloor \cI d(\lambda))=\sum\limits^{2r+1}_{|J|=0}(-1)^{|J|}d_J(\psi^\rho\frac{\partial E_\rho(\lambda)}{\partial y^\sigma_J})\omega^\sigma \wedge ds
 =
\eEq
\bEq\label{eq:coordJacobi}
 =\sum\limits^{2r+1}_{|J|=0} d_J\psi^\sigma\frac{\partial E_\rho (\lambda)}{\partial y^\sigma_J}\omega^\rho\wedge ds 
\,.
\eEq
In the following we use the notation $\mathcal{J}_{\psi}(\lambda)$ for short to denote  $ \cI d(\psi\rfloor \cI d(\lambda))$.

\bDf
Let $\lambda$ be a Lagrangian of order $r$. A {\em Jacobi field} for the Lagrangian $\lambda$ is a vertical vector field $\psi$ that belongs to the kernel of the Jacobi morphism, \ie satisfying the {\em Jacobi equation} for the Lagrangian $\lambda$
\beq 
\mathcal{J}_{\psi}(\lambda)=0 \,.
\eeq 
\eDf

\bRm
The Jacobi morphism $\mathcal{J}_\psi(\lambda)$, evaluated along an extremal $\gamma$, depends only on the values of the vector field $\psi$ along $\gamma$.

A Jacobi equation along an extremal is then well defined; we call its solutions the {\em Jacobi fields along an extremal} $\gamma$.
\eRm 

\bRm
Equation \eqref{eq:coordJacobi} provides the `adjoint expression' for the Jacobi equation along extremals; it can be of use in order to obtain an easier characterization of the kernel of the Jacobi morphism in practical computations.

Notably here it will be used in order to calculate the conserved current associated with invariant Yang--Mills connections (see {\em Example \ref{Jac_conservation_ex}} below).

See also \cite{PaWi19} for an explicit application in $SU(3)$-Yang--Mills theories in the context of a  variationally featured symmetry breaking in view of a canonical characterization of confinement phases in non-abelian gauge theories \cite{'tHo81}.
\eRm

\subsection{Jacobi fields and higher conservation laws}

We observe that the Jacobi equation for variations of $h\rho$ can be expressed in terms of $h\rho$. In fact, just using the exactness of the representation sequence, we have
\begin{equation*}
\begin{aligned}
\cI d(\psi\rfloor \cI d(\rfloor L_{J^{r+1}\psi_s}&\dots L_{J^{r+1}\psi_1}h\rho))=\cI d(\psi\rfloor \cI d(\psi_s \rfloor \cI d(\dots \psi_1\rfloor \cI d(h\rho)\dots))) \,.
\end{aligned}
\end{equation*}
The application of Theorem \ref{Jacobicons} to an iterated variation of a Lagrangian give results that are relevant for the Lagrangian itself; in fact, using \eqref{eq:highvar} we can relate the Noether current of the $s$-th variation with Noether currents of lower order variations. More precisely, we can state the following important {\em original} result.

\bTh\label{jacobivar}
If we take a symmetry of an $(l-1)$-th variation of $h\rho$ and we suppose that the $s$-th variation ($s<l$) is taken with respect to a Jacobi field of the $(s-1)$-th variation, then
\begin{equation*}
d_H\epsilon_{\psi_l}\dots d_H\epsilon_{\psi_{s+1}}(L_{J^{r+1}\psi_s}\dots L_{J^{r+1}\psi_1}h\rho)=0 \,.
\end{equation*}
\eTh
\bPf
Actually we have
\beq d_H\epsilon_{\psi_l}(L_{J^{r+1}\psi_{l-1}}\dots L_{J^{r+1}\psi_{1}}h\rho)=0 \,,
\eeq 
with some terms that vanish separately. In fact, applying the definition of Jacobi field, we get
\begin{equation*}
\begin{aligned}
\psi_{l}\rfloor \cI d(\psi_{l-1}\rfloor \cI d(\dots \psi_{2} \rfloor \cI d(\psi_{1}\rfloor \cI d(h\rho))\dots)=0 \,,\\
d_H\epsilon_{\psi_l}(\psi_{l-1}\rfloor \cI d(\dots \psi_{2} \rfloor \cI d(\psi_{1}\rfloor \cI d(h\rho))\dots)=0 \,,\\
\dots \\
d_H\epsilon_{\psi_l} \dots d_H \epsilon_{\psi_{s+2}}(\psi_{s+1}\rfloor \cI d(\dots(\psi_1\rfloor \cI d(h\rho))\dots )=0 \,.
\end{aligned}
\end{equation*}
 Then the statement follows easily using \eqref{eq:highvar}  (Theorem \ref{th:highvar}).
\ePf
\bRm
The previous result is a strong conservation law(\ie it holds along any section, not necessarily an extremal) : the conserved current is the $(l-s-1)$-th variation of the horizontal differential of the Noether current for the $s$-th variation of $h\rho$. The result is not trivial because we are not assuming that $\psi_{s+1}$ is a symmetry of the $s$-th variation.
\eRm

\bRm \label{Higher Jacobi}
We can write formula \eqref{eq:highvar} in terms of Jacobi morphisms:
\begin{equation*}
\begin{aligned}
(\pi_{r_l,r+1})^*(L_{J^{r+1}\psi_l}\dots &L_{J^{r+1}\psi_1}h\rho)=\\
=&\psi_{l,V}\rfloor \mathcal{J}_{\psi_{l-1,V}}(\psi_{l-2,V}\rfloor\mathcal{J}_{\psi_{l-3,V}}\dots (h\rho)\dots) +\\
+&d_H\epsilon_{\psi_l}(\psi_{l-1,V}\rfloor \mathcal{J}_{\psi_{l-2,V}}(\psi_{l-3,V}\rfloor \dots (h\rho)\dots )) +\\
+&d_H\epsilon_{\psi_l}(d_H\epsilon_{\psi_{l-1}}(\psi_{l-2,V}\rfloor \mathcal{J}_{\psi_{l-3,V}}(\dots(h\rho)\dots))) +\\
+&\dots+d_H\epsilon_{\psi_l}(d_H\epsilon_{\psi_{l-1}}d_H(\dots d_H\epsilon_{\psi_1}(h\rho)\dots)) \,.
\end{aligned}
\end{equation*}
\eRm

\bRm \label{REMARK}
Note that by Theorem \ref{MAIN}, and by Theorem \ref{th:Jacobi}, Jacobi fields along extremals are vertical symmetry transformations of extremals and {\em vice versa}.
\eRm

Our characterization of (vertical) symmetry transformations of extremals as Jacobi fields along extremals is motivated by the fact that there  can exist  conservation laws associated with symmetry transformations, which in principle are different from the Noether or the Noether--Bessel-Hagen conservation laws associated with symmetries or generalized symmetries of a Lagrangian $\lambda$, in particular we stress once more that {\em all symmetries of equations are also symmetry transformations of extremals, but the converse is not true in general}.

Indeed, we have the following our {\em further main result}.

\bTh \label{Jacobicons}
Let $\rho$ be an $n$-form on $J^{r-1}Y$ and $h\rho$ the associated Lagrangian on $J^rY$.
Let $\psi_1$ and $\psi_2$ on $Y$ be two generators of  vertical symmetry transformations of extremals. 
Then, along
extremals of $h\rho$,  the  weak conservation law holds true:
\bEq\label{eq:Jacobicons2}
d_H\epsilon_{\psi_2}(\psi_1\rfloor \cI d(h\rho))=0 \,.
\eEq
\eTh

\bPf  
Indeed, by Theorem \ref{MAIN}, the two generators of symmetry transformations $\psi_1$ and $\psi_2$ are also Jacobi fields, \ie  they must satisfy 
$\mathcal{J}_{\psi_i}(h\rho) = 0$, for $i=1,2$. Therefore from \eqref{eq:2nd_commu}, since also $[\psi_2,\psi_1]\rfloor \cI d (h\rho)$ vanishes along extremals, we get the result.
\ePf

For the interpretation of this current as a Noether current for a `deformed' Lagrangian, see the note at the end of Proposition \ref{th:conslaws}.
\bRm
Suppose that $\psi_2$ is a symmetry of the first variation of $h\rho$ generated by $\psi_1$ and that $\psi_1$ and $\psi_2$ satisfy
$\psi_2\rfloor\mathcal{J}_{\psi_1}(h\rho)=0$,
then we have a strong conservation law:
\bEq\label{eq:Jacobicons1}
d_H\epsilon_{\psi_2}(L_{J^{r}\psi_1 }h\rho)=0 \,.
\eEq
 Now, it is clear that {\em along extremals}, taking two vertical symmetry transformations $\psi_1$ and $\psi_2$, we get two separated weak conservation laws; see also \cite{AcPa20}.
\eRm

\section{Conserved currents for invariant Yang--Mills connections} \label{Jac_conservation}

Let  us consider a Yang--Mills theory \cite{YaMi54} on the bundle $(C_P, \pi ,M)$ of principal connections with structure bundle $(P, p, M, G)$,  $G$ being a {\em semi-simple group}. 
Lower Greek indices label space-time coordinates, while capital Latin indices label the Lie algebra $\mathfrak{g}$ of $G$, then, on the bundle $C_P$, we introduce coordinates $(x^\mu, \omega^A_\sigma)$.

Let  $\delta$ be the Cartan-Killing metric on the Lie algebra $\mathfrak{g}$, and choose a $\delta$-orthonormal basis $T_A$ in $\mathfrak{g}$; the components of $\delta$ will be denoted $\delta_{AB}$. 
The Yang-Mills Lagrangian is locally expressed by
\beq 
\lambda_{YM}=-\frac{1}{4}F^A_{\mu \nu}g^{\mu \rho}g^{\nu \sigma}F^B_{\rho \sigma}\delta_{AB}\sqrt{g}ds \,,
\eeq 
where $g$ stands for the absolute value of the determinant of the metric $g_{\mu \nu}$, 
 $c^A_{BC}$ are the {\em structure constants} of $\mathfrak{g}$, 
$F^A_{\mu \nu}=\omega^A_{\nu,\mu}-\omega^A_{\mu,\nu}+c^A_{BC}\omega^B_\mu\omega^C_\nu$ is the so called  {\em field strength}, and we set $\omega^A_{\mu,\nu}=d_\nu\omega^A_\mu$.

From now on, we assume the metric $\eta$ to be Minkowskian; in this case the Euler--Lagrange expressions for the Yang--Mills Lagrangian are explicitly written as 
 \bEq\label{Euler}
E^\nu_B = \delta_{BA}\eta^{\lambda \mu}\eta^{\epsilon \nu}(\omega^A_{\epsilon, \lambda \mu}-\omega^A_{\lambda, \epsilon \mu}+c^A_{ZD}\omega^Z_{\lambda,\mu}\omega^D_\epsilon+c^A_{ZD}\omega^Z_\lambda\omega^D_{\epsilon,\mu})
 + \\
 + \eta^{\lambda\mu}\eta^{\epsilon\nu}\delta_{DA}(\omega^D_{\epsilon,\lambda}-\omega^D_{\lambda,\epsilon}+c^D_{EF}\omega^E_\lambda\omega^F_\epsilon)c^A_{BZ}\omega^Z_\mu \,. \nonumber
\eEq

Let $(\phi^a)$  be a set of coordinates on the group $G$. A vertical vector field over $C_P$ has the form $\psi=\psi^Z_\sigma \frac{\partial}{\partial \omega^Z_\sigma}$ and its components 
satisfy the transformation rule
$ \psi'^B_\nu=Ad^B_A(\phi)\psi^A_\mu\overline{J}^\mu_\nu$
where $Ad^B_A(\phi)$ is the adjoint representation of $G$ on $\mathfrak{g}$ and $\overline{J}^\mu_\nu$ denotes the inverse of the matrix of the change of coordinates in the base space.
Let $L(M)$ be the frame bundle of $M$, $V=\mathfrak{g}\otimes\mathbb{R}^n$.
Let us denote by $\nabla$ the covariant derivative corresponding to $\Omega = dx^\mu\otimes(\partial_\mu+ c^B_{AD}\psi^D_\sigma\omega^A_\mu\partial^{\sigma}_B) $ the connection induced on 
the bundle $(P\times_M L(M))\times_\lambda V$ of vertical vector fields over $C_P$, where the representation $\lambda$ is comes from by the transformation rules above (see \cite{Kol79,KMS93,FFP01}).

By some careful  manipulations (see \cite{AcPa20} for details), Theorem \ref{th:Jacobi} and specifically formula \eqref{eq:coordJacobi} provide  {\em the Jacobi equation along extremals} for this kind of Yang-Mills theory. In particular, due to the antisymmetry of $F^D_{\beta\sigma}$ in the lower indices, it splits in the antisymmetric and symmetric parts
\bEq\label{Equazione Jacobi}
\eta^{ \nu[\sigma }\eta^{\beta] \alpha }\left\{
\nabla_\beta\left[\left(\nabla_\alpha\psi^A_\sigma-\nabla_\sigma\psi^A_\alpha\right)\delta_{BA}\right]+
F^D_{\beta\sigma}\delta_{AD}c^A_{BZ}\psi^Z_\alpha\right\}=0 \,,
\\
\eta^{ \nu(\sigma }\eta^{\beta) \alpha }\left\{
\nabla_\beta\left[\left(\nabla_\alpha\psi^A_\sigma-\nabla_\sigma\psi^A_\alpha\right)\delta_{BA}\right] 
\right\}=0 \,. \nonumber 
\eEq
for any pair $(\nu,B)$; hereafter the brackets $( )$ and $[ ]$ in the superscripts denote symmetrization and anti-symmetrization, respectively.

Note that the left hand side of these equations are the analogous, for a Minkowskian metric, of the classical expression for the Jacobi operator for Yang--Mills theories on different backgrounds, see \eg  \cite{AtBo83,Bou87}, and it reproduces results for first order non regular Lagrangians \cite{GoSt73}.

In order to avoid notational confusion, let  $\chi^A_\mu, \chi^A_{\mu,\nu}, \chi^A_{\mu,\nu\rho}, \dots$ denote generators of contact forms.
\bRm
According with our results, the solutions  $\psi$ of the above equations are the generators of symmetry transformations of Yang--Mills extremals $\omega$,  \ie if $\psi$ is a solution of the above equation, then the source form $L_{J^{3}\psi} (E^\nu_B  \,\chi_\nu^B \wedge ds )$, where $E^\nu_B$ are given by \eqref{Euler}, also vanishes along the same extremals, \ie 
\beq
(L_{J^{3}\psi} 
(E^\nu_B\,\, \chi_\nu^B \wedge ds ) )\circ  J^{3}\omega =0 \,.
\eeq
Here, by a slight abuse of notation, we denoted by $\omega$ a section  of the bundle $(C_P, \pi ,M)$ which is an extremal, \ie a Yang--Mills connection.

As we already  mentioned, compared with transformations leaving invariant the Euler--Lagrange form $E^\nu_B  \,\chi_\nu^B \wedge ds$, such transformations are involved with a weaker  invariance property, since the Euler--Lagrange form is not invariant under their action, but it is transformed to a source form having the same extremals.  
Note that indeed the Yang--Mills extremals $\omega$ are also solutions of the equation above {\em and vice versa}.
\eRm

\bEx \label{Jac_conservation_ex}
As an illustration of the application of our main result,  Theorem \ref{Jacobicons}, we determine  now the conserved current associated to such a weaker invariance property. 
We write down explicitly the current for two given generators of vertical symmetry transformations $\psi$ and $\tilde{\psi}$, solutions of  equation \eqref{Equazione Jacobi} (for details we refer to \cite{Ac17,AcPa20}, where computations are made for Jacobi fields along extremals). 

Being the vector fields vertical, from Theorem \ref{Jacobicons}, equation \eqref{eq:Jacobicons2}, and according to Theorem \ref{var_lie_der}, the conserved current  along an extremal is defined through $p_1\mathcal{R}(d(\psi\rfloor \cI d(\lambda_{YM}))$.
Recalling that $E^\nu_B$ are coordinate expression of the Euler--Lagrange form according to \eqref{Euler}, we apply the coordinate characterization of the residual operator (given in Example \ref{residual_operator}) to the form
\beq
& d (\psi\rfloor \cI d (\lambda_{YM}))
  = (\frac{\partial \psi^B_\nu}{\partial \omega^Z_\rho}E^\nu_B+\psi^B_\nu\frac{\partial E^\nu_B}{\partial \omega^Z_\rho})\chi^Z_\rho\wedge ds+ \\ & + (\psi^B_\nu\frac{\partial E^\nu_B}{\partial \omega^Z_{\rho,\xi}})\chi^Z_{\rho,\xi}\wedge ds+(\psi^B_\nu\frac{\partial E^\nu_B}{\partial \omega^Z_{\rho,\xi\tau}})\chi^Z_{\rho,\xi\tau}\wedge ds \,.
\eeq
By suitably rewriting the above in the form $\sum\limits^2_{|I|=0}d_I(
\chi^A_\mu\wedge\zeta^{\mu,I}_A)$, 
we can easily obtain the local expression for 
$\mathcal{R}(d(\psi\rfloor \cI d(\lambda_{YM})))$, 
thus obtaining the conserved current we are looking for:
\beq
\epsilon_{\tilde{\psi}}(\psi\rfloor \cI d(\lambda_{YM}))&  = & [\eta^{\rho[\xi}\eta^{\sigma]\nu}\delta_{BA}c^A_{ZD}\omega^D_\sigma(\psi^B_\nu\tilde{\psi}^Z_\rho-\psi^Z_\rho\tilde\psi^B_\nu)+  \\
& & (\eta^{\xi\sigma}\eta^{\rho\nu}-\eta^{\rho(\sigma}\eta^{\xi)\nu})(\psi^Z_\nu\nabla_\sigma(\tilde\psi^B_\rho\delta_{BZ})-\tilde{\psi}^Z_\rho\nabla_\sigma(\psi^B_\nu\delta_{BZ}))]ds_\xi \,.
\eeq

\bRm
It is noteworthy that, by  Proposition \ref{th:conslaws} and in particular by Remark \ref{REMARK}, here the existence and the meaning of the above current (also appeared in \cite{AcPa20} in relation with Jacobi fields) is understood under a new light, definitely {\em relevant from a physical point of view}.

In the present paper we clarify that such a conservation law emerges by an invariance property of the set of extremals and, moreover, that the associated conserved current can be interpreted as a very specific kind of Noether current, the existence of which is related with a wide class of symmetry transformations.
Indeed, we proved that this current can be identified as the  Noether current for the Yang--Mills Lagrangian `deformed'  by the symmetry transformation of extremals $\psi$  and associated with (or generated by) the symmetry transformation of extremals $\tilde\psi$.
\eRm

\bRm
We note that Equation \eqref{eq:2nd_commu} of Proposition \ref{th:conslaws} says us that for any vertical vector field $\psi_1=\psi_2=\zeta$, the current $\epsilon_{\zeta}(\zeta\rfloor \cI d(h\rho))$ is a strong conserved current (\ie conserved  `off shell'). 
However, it can be easily checked that, at least in the specific case of study, for any (vertical) symmetry transformation of extremals $\tilde{\psi} = \psi$ the weak  (\ie `on shell') conserved current  reduces to 
$\eta^{\rho(\sigma}\eta^{\xi)\nu}(\psi^Z_\nu\nabla_\sigma(\psi^B_\rho\delta_{BZ})- \psi^Z_\rho\nabla_\sigma(\psi^B_\nu\delta_{BZ})) ds_\xi $, which vanishes identically  because 
$\eta^{\rho(\sigma}\eta^{\xi)\nu} = \eta^{\nu(\sigma}\eta^{\xi)\rho}$. This holds true for any couple of linearly dependent symmetry transformations.
\eRm

\eEx

\section*{Acknowledgements}
Research partially supported by Department of Mathematics - University of Torino through the  projects FERM$\_$RILO$\_17\_ 01$ and PALM$\_$RILO$\_20\_ 01$ (MP) and written under the auspices of GNSAGA-INdAM. The first author (LA) is also supported by a NWO-UGC project, Grant BM$.00193.1$.
The second author (MP) would like to acknowledge the contribution of the Cost Action CA17139 - European Topology Interdisciplinary Action.


\end{document}